\documentclass[prb, preprint, amsmath,amssymb]{revtex4-1}

\usepackage{graphicx}
\usepackage{bm}
\usepackage{color}

\def\etal{{\it et~al.}}

\def\o{\omega}
\def\s1{\sigma_1(\omega)}

\begin{document}

\title{\boldmath Infrared ellipsometry study of the confined electrons in a high-mobility $\gamma$-Al$_2$O$_3$/SrTiO$_3$ heterostructure\unboldmath}

\author{M. Yazdi-Rizi}%
\email{meghdad.yazdi@unifr.ch}
 \affiliation{University of Fribourg, Department of Physics and Fribourg Center for Nanomaterials,
Chemin du Mus\'{e}e 3, CH-1700 Fribourg, Switzerland}
\author{P. Marsik}
\affiliation{University of Fribourg, Department of Physics and Fribourg Center for Nanomaterials,
Chemin du Mus\'{e}e 3, CH-1700 Fribourg, Switzerland}
\author{B. P. P. Mallett}
\affiliation{University of Fribourg, Department of Physics and Fribourg Center for Nanomaterials,
Chemin du Mus\'{e}e 3, CH-1700 Fribourg, Switzerland}
\author{A. Dubroka}
\affiliation{Department of Condensed Matter Physics, Faculty of Science and Central European Institute of Technology, 10 Masaryk University, Kotl\'{a}\v{r}ska 2, 61137 Brno, Czech Republic}
\author{D. V. Christensen}
\affiliation{Department of Energy Conversion and Storage, Technical University of Denmark, Ris{\o} Campus, 4000 Roskilde, Denmark}
\author{Y. Z. Chen}
\affiliation{Department of Energy Conversion and Storage, Technical University of Denmark, Ris{\o} Campus, 4000 Roskilde, Denmark}
\author{N. Pryds}
\affiliation{Department of Energy Conversion and Storage, Technical University of Denmark, Ris{\o} Campus, 4000 Roskilde, Denmark}
\author{C. Bernhard}%
\email{christian.bernhard@unifr.ch}
\affiliation{University of Fribourg, Department of Physics and Fribourg Center for Nanomaterials,
Chemin du Mus\'{e}e 3, CH-1700 Fribourg, Switzerland\\}

\date{\today}

\begin{abstract}

With infrared ellipsometry we studied the response of the confined electrons in a $\gamma$-Al$_2$O$_3$/SrTiO$_3$ heterostructure in which they originate predominantly from oxygen vacancies. From the analysis of a so-called Berreman mode, that develops near the highest longitudinal optical phonon mode of SrTiO$_3$, we derive the sheet carrier density, ${N_{s}}$, the mobility, $\mu$, and also the depth profile of the carrier concentration. Notably, we find that \textit{N$_{s}$} and the shape of the depth profile are similar as in LaAlO$_3$/SrTiO$_3$ heterostructures for which the itinerant carriers are believed to arise from a polar discontinuity. The main differences concern the higher mobility and a relatively stronger confinement of the electrons in $\gamma$-Al$_2$O$_3$/SrTiO$_3$.

\end{abstract}



\maketitle


\section{Introduction}
The observation that highly mobile electrons can be created at the interface between the nominal band-insulators SrTiO$_3$ (STO) and LaAlO$_3$ (LAO) with $\Delta_\mathrm{gap}^\mathrm{STO} = 3.2$ eV and $\Delta_\mathrm{gap}^\mathrm{LAO}=5.6$ eV, respectively, has initiated intense research activities on LAO/STO heterostructure~\cite{Mannhart2008}. Meanwhile, it has been demonstrated that functional field effect devices can be made from these oxide heterostructures~\cite{Cen2009} which even allow one to tune a superconductor to insulator quantum phase transition at very low temperatures~\cite{Caviglia2008}. A major unresolved question concerns the origin of these confined charge carriers and the conditions for obtaining a high mobility. The explanations range from (i) an electronic reconstruction due to the polar discontinuity at the LAO/STO interface which gives rise to a transfer of 1/2 e charge per STO unit cell (corresponding to a sheet carrier density of \textit{N$_{s}$}=3.3$\times10^{14}$ cm$^{-2}$), (ii) an intermixing of La and Sr cations across the interface~\cite{Willmott2007}, to (iii) the creation of oxygen vacancies which may be stabilized near the interface by space charge effects~\cite{Kalabukhov2007}.
While the polar discontinuity scenario (i) has obtained a great deal of attention, more recently the focus has shifted toward the oxygen vacancies (iii). In particular, it has been demonstrated that an electron gas can be created in the absence of a polar discontinuity, for example at the LAO/STO (110) interface~\cite{Herranz2012}, with amorphous LAO cap layers~\cite{Chen2011, Lee2012}, and even with a $\gamma$-Al$_2$O$_3$ (GAO) top layer which has a spinel structure. Note that the latter remains polar, but with a much reduced polarity compared to LAO/STO~\cite{Chen2013, Kormondy2015}. The results reported in Ref.~\cite{Liu 2013} suggest that both the oxygen vacancies and the polar discontinuity can be the source of mobile electrons that are confined near the surface of STO. Notably, the highest mobility reported to date has been achieved in GAO/STO heterostructures for which there is clear evidence that oxygen vacancies are at the heart of the interfacial electrons~\cite{Chen2013, Kormondy2015}.
 
This raises the question about the differences in the dynamical properties of the interfacial electrons and also of the concentration depth profiles in these different kinds of heterostructures. In GAO/STO the latter has so far only been determined at room temperature using x-ray photoelectron spectroscopy (XPS)~\cite{Chen2013, Schutz2015}. The depth distribution of the confined electrons at low temperature, where the dielectric constant of STO is much larger and the electron mobility strongly enhanced, is yet unknown. Some of us have previously shown for the case of LAO/STO that this information can be obtained with infrared ellipsometry~\cite{Dubroka2010}. Here we present a corresponding ellipsometry study of a GAO/STO heterostructure.

\section{Sample growth and characterization}

The GAO thin film with a thickness of 2.5 unit cells has been grown by pulsed laser deposition (PLD) on a singly TiO$_2$-terminated (001) STO crystal with a repetition rate of 1 Hz and a laser fluence of 2.5 J cm$^{-2}$, and monitored by in-situ high pressure RHEED~\cite{Chen2013}. During the deposition at 600~$^{\circ}$C, the oxygen pressure was kept constant at 10$^{-4}$ mbar. After the deposition, the sample was cooled down to room temperature at the deposition pressure without any further post oxygen annealing. The sample has been studied with standard dc Hall transport measurements which yield typical values for the sheet carrier density and mobility of \textit{N$_{s}$}$\approx$ 6.0$\times10^{13}$ cm$^{-2}$ and $\mu\approx11000$ cm$^{2}$/Vs at 2 K and \textit{N$_{s}$}$\approx5.7\times$10$^{13}$ cm$^{-2}$ and $\mu\approx$ 9700 cm$^{2}$/Vs at 10K.  

Infrared ellipsometry measurements have been performed with a home-built setup that is equipped with a He flow cryostat and attached to a Bruker 113V Fast Fourier spectrometer as described in Ref.~\cite{Bernhard2004}. The data have been taken at 10K in rotating analyser mode, with and without a static compensator based on a ZnSe prism. Very similar data (not shown) have been obtained by using a rotating compensator with a ZnSe prism. The angle of incidence of the light was set to 75 degree. Special care was taken to avoid photo-doping effects by shielding the sample against visible and UV light~\cite{Kozuka2007}. The measurements have been performed under identical conditions first on the GAO/STO heterostructure and, right afterwards, on a bare STO substrate that serves as a reference.

\section{Experimental Results}

Infrared ellipsometric spectra contain valuable information about the properties of the electrons that are confined at the interface of the GAO/STO heterostructure. The most instructive feature is due to a so called Berreman mode that occurs in the vicinity of the plasma frequency of the highest longitudinal optical (LO) phonon mode~\cite{Berreman1963, Humlíček1996} which in STO is located at $\omega_{LO}$(STO)$\approx788$ cm$^{-1}$ ~\cite{Rossle2013}. It is an electronic plasma mode that originates from a charging of the interfaces due to the dynamical accumulation of the itinerant charge carriers. It occurs under an oblique angle of incidence of the light if the polarization of the electric field, $E$, is parallel to the plane of incidence (\textit{p}-polarization) and thus has a finite normal component with respect to the interfaces. The corresponding reflection coefficient, \textit{r$_{p}$}, exhibits a characteristic structure in the vicinity of the LO plasma frequency,  $\omega_{LO}$. The Berreman mode is conveniently presented and analyzed in terms of the difference spectrum of the ellipsometric angle, $\Psi$=arctan(\textit{r$_{p}$}/\textit{r$_{s}$}), of the heterostructure with respect to the bare STO substrate, $\Delta\Psi$=$\Psi$(sample)-$\Psi$(STO).

The spectrum of $\Delta\Psi$ shown in Figure 1(a) reveals two major features. The broad peak with a maximum near 956 cm$^{-1}$ corresponds to the Berreman-mode. It has been previously pointed out that the difference in frequency between this maximum and $\omega_{LO}$(STO) is a measure of the plasma frequency of the itinerant electrons, $\omega_{pl}$~\cite{Dubroka2010}. The intensity of this peak is determined by the overall sheet carrier density and its broadening by the mobility (or the inverse scattering rate). For the case of LAO/STO it was shown that this peak has a strongly asymmetric shape which provides additional information about the depth-profile of the carrier concentration~\cite{Dubroka2010}. For the LAO/STO system the carrier density is highest next to the interface and decreases rather rapidly toward the bulk of STO over a length scale of about 11 nm.

The second feature is a fairly sharp and pronounced minimum around $\omega^{dip}\approx$865 cm$^{-1}$ that occurs slightly above $\omega_{LO}$(STO) where the real part of the dielectric function of STO matches the one of the GAO top layer such that the interface becomes fully transparent. As was outlined in Ref.~\cite{Skoromets2014} and Ref.~\cite{Dubroka2010}, this dip feature contains contributions from the out of plane and the in plane components of the dielectric function. The in plane contribution arises from the Drude-like response of the itinerant electrons which leads to a reduction of $\varepsilon_{1}$ (865 cm$^{-1}$) and, given a low mobility and thus large scattering rate, an increase of $\varepsilon_{2}$ (865 cm$^{-1}$). The strength of this dip is a measure of the overall carrier density \textit{N$_{s}$}, it is rather insensitive to the details of the depth distribution of the itinerant carriers (since the penetration depth of the infrared light is on the order of a micrometer). It is also not sensitive to the in plane mobility of the carriers, unless it is very low such that the Drude peak in $\varepsilon_{2}$ is very broad and extends to the dip feature.

We start by analyzing the ellipsometric spectra of GAO/STO with the same model that was used in Ref.~\cite{Dubroka2010} for the LAO/STO heterostructures. Figure 1(a) shows the measured spectrum of $\Delta\Psi$ (symbols) at 10K and the best fits using either a block like potential with a constant (green solid line) or a graded profile of the concentration of the confined carriers (red solid line). The graded profile provides a significantly better fit to the data. It reproduces the main features, like the pronounced dip around 865 cm$^{-1}$, the maximum around 955 cm$^{-1}$ and, especially, the long tail toward higher frequency which is terminated by a step like feature around 1186 cm$^{-1}$. Figure 1(b) shows a comparison with the data and the best fit with the graded profile for the LAO/STO heterostructure~\cite{Dubroka2010}. The obtained depth profiles for GAO/STO (blue solid symbols) and the LAO/STO (red solid symbols) heterostructures are displayed in Figure 1(c). The derived values for the sheet carrier density \textit{N$_{s}$}, the mobility, $\mu$, and the total thickness of the conducting layers, \textit{d}, are listed in Table 1. To allow for a direct comparison, we assumed the same value of the effective mass, \textit{m*}=3.2 \textit{m$_{e}$}, (\textit{m$_{e}$} is the free electron mass) as in Ref.~\cite{Dubroka2010}. First of all, we notice that the obtained sheet carrier density of 
$\textit{N$_{s}$}(IR)\approx$ 6.2$\times$10$^{13}$ cm$^{-2}$ is rather close to the one of the dc Hall transport measurements of $\textit{N$_{s}$}(Hall)\approx$ 5.7$\times$10$^{13}$ cm$^{-2}$. Secondly, we remark that the larger thickness of the conducting layer of \textit{d}=7.5 nm, as compared to the one deduced from XPS measurements at room temperature of \textit{d}=0.9 nm~\cite{Chen2013, Schutz2015}, can be understood in terms of the much larger dielectric constant of STO at low temperature which leads to enhanced screening and thus a weaker confinement of the electrons. A similar difference between the thicknesses deduced from the infrared measurements at 10K~\cite{Dubroka2010} and XPS~\cite{Sing2009} and scanning probe measurement at room temperature~\cite{Basletic2008} was previously observed for LAO/STO. Overall, it is rather striking that despite the supposedly different origin of the confined electrons, oxygen vacancies in GAO/STO and the polar discontinuity in LAO/STO, respectively, the value of \textit{N$_{s}$} and the shape of the depth profile are very similar for both systems. The major difference concerns the mobility of the charge carriers which is significantly larger in GAO/STO than in LAO/STO and the reduced thickness, \textit{d}, which suggests a stronger confinement of the carriers. The higher mobility enhances the visibility of the characteristic features of the Berreman mode such as the peak and, especially, the step-like edge that terminates the high energy tail at 1185 cm$^{-1}$. The latter is a measure of the maximal plasma frequency next to the interface (see Figure 1(c)). Note that the obtained mobility is still much lower than the values reported from dc transport measurements~\cite{Chen2013}. A similar difference was observed for LAO/STO where it was explained in terms of the different frequency scales that are probed by the dc transport and the optical experiment. The charge carriers in STO have indeed a polaronic character~\cite{van Mechelen2008} and thus are subject to inelastic scattering with the lattice that becomes very pronounced in the frequency range of the Berreman mode. 

While this isotropic model accounts for the main features of the Berreman-mode, it fails to describe some details of the $\Delta\Psi$ curve. In particular, it overestimates the depth of the dip feature at 865 cm$^{-1}$. In the following we show that the fitting can be further improved by allowing for an anisotropic mobility of the charge carriers along the out of plane and the in plane directions, i.e. $\mu_\textit{z}\neq\mu_\textit{xy}$. It is well known that the confined electrons at the STO interface may originate from different bands derived from the \textit{t$_{2g}$} orbitals which have an anisotropic dispersion behavior. As was discussed in the previous paragraph, the peak of the Berreman mode is governed by the out of plane component of the dielectric function to which the electrons in the \textit{d$_{xz}$}- and \textit{d$_{yz}$}- related bands provide the major contribution since they are dispersive along the \textit{z}-direction. The contribution of the \textit{d$_{xy}$}-related bands is expected to be less important, since their \textit{z}-axis conductivity arises mainly from transitions between the different subbands~\cite{Rossle2013}. On the other hand, it can be expected that the electrons with \textit{d$_{xy}$} character govern the Drude-like response along the in plane direction which makes a major contribution to the dip feature at 865 cm$^{-1}$.

Figure 2 (a) shows the best fit with an anisotropic Drude-model, for which the mobility along the in-plane direction, $\mu_\textit{xy}$, was allowed to vary, whereas $\mu_\textit{z}$ and \textit{N$_{s}$} as well as the depth profile where kept fixed at the values obtained with the isotropic model. For comparison, the red solid line in Figure 2(b) shows the sole contribution of the \textit{z}-component, which has been singled out by setting $\mu_\textit{xy}=0$ cm$^{2}/Vs$. It confirms that the Berreman-mode, i.e. the region above the dip where $\Delta\Psi>0$, is well accounted for by the \textit{z}-axis component. The anisotropic model provides indeed a better description of the dip feature than the graded isotropic model in Figure 1(a). However, it yields an unreasonably low value of $\mu_\textit{xy}=5$ cm$^{2}$/Vs in view of the very high mobility that has been deduced from the dc measurements in Ref.~\cite{Chen2013}.
We therefore allowed for two Drude components to account independently for the out of plane response (with \textit{N$_{s,z}$}, $\mu_\textit{z}>0$ and $\mu_\textit{xy}=0$) and the in plane response (with \textit{N$_{s,xy}$}, $\mu_\textit{xy}>0$ and $\mu_\textit{z}$=0), respectively. The parameters \textit{N$_{s,xy}$} and $\mu_\textit{xy}$ were allowed to vary, whereas \textit{N$_{s,z}$} and $\mu_\textit{z}$ as well as the depth profile of the carrier concentration were fixed to the ones obtained from the isotropic model. Note that the fitting is not sensitive to the depth profile of the in plane component which mainly contributes to the dip feature but hardly to the peak in $\Delta\Psi$. Therefore, we used for the in-plane component a block-like depth profile with a thickness of 7.5 nm. The best fit shown in Figure 2(c) yields a reduced value of \textit{N$_{s}$}=4.6$\times$10$^{13}$ and a larger mobility of $\mu_\textit{xy}>30$ cm$^{2}$/Vs. According to the discussion in the previous paragraph, the fitting only yields a lower limit for the in plane mobility since it looses sensitivity when the Drude peak becomes too narrow. The fit in Figure 2(c) provides a satisfactory description of the experimental data that could not be significantly improved by allowing additional parameters to vary. A corresponding anisotropic fitting procedure for the LAO/STO heterostructure is not reported here since the broadening of the characteristic features of the Beremann mode, due to the significantly lower electron mobility of LAO/STO, makes the distinction between the out of plane and the in plane components rather difficult and unreliable.

We have thus obtained the following information about the high mobility electrons in GAO/STO. From the analysis of the Berreman mode in the range where $\Delta\Psi>0$, we derived the parameters of the \textit{z}-axis component of the response of the confined carriers with \textit{N$_{s,z}$}=6.2$\times$10$^{13}$ cm$^{-2}$ and $\mu_\textit{z}$=75 cm$^{-2}$/Vs and determined the shape of the graded depth profile of the carrier concentration which has a thickness of \textit{d}=7.5 nm. In comparison, we have obtained only limited information about the in plane response. The latter is mostly based on the analysis of dip feature around 865 cm$^{-1}$ and yields \textit{N$_{s,xy}$}=4.6$\times$10$^{13}$ cm$^{-2}$ and a lower limit for the in plane mobility of  $\mu_\textit{xy}>30$ cm$^{-2}$/Vs. The width and the shape of the depth profile of the carriers that are mobile in the lateral direction along the interface could not be determined.  
The comparison with the LAO/STO heterostructures thus yields very similar values of \textit{N$_{s}$} and the same characteristic, asymmetric shape of the depth profile. The main difference concerns the significantly higher mobility of the confined electrons and the reduced thickness of their distribution near the interface, although the GAO/STO is non-annealed. The higher mobility in GAO/STO may be understood in terms of the perfect lattice match thus much reduced strain that is imposed on the interfacial region of STO by the GAO top layer and, accordingly, a lower concentration and different distribution of the structural defects~\cite{Chen2013, Kormondy2015}. The reduced thickness of the conducting layer likely reflects a larger magnitude of the confining potential (that is consistent with the slightly larger value of \textit{N$_{s}$}) rather than a reduced strength of the dielectric screening in STO. The latter would require a hardening of the soft mode in STO that results from defects and compressive strain which both are expected to be stronger in LAO/STO than in GAO/STO.

\section{Summary}

With infrared ellipsometry we have studied the so-called Beremann mode in GAO/STO heterostructures which arises from the itinerant charge carriers that are confined to the interface. We analyzed the Beremann mode with the same model that was previously reported in Ref.~\cite{Dubroka2010} for LAO/STO heterostructures and found that both samples have a similar sheet carrier densities and asymmetric shapes of the depth profile of the carrier concentration. This is rather surprising since the main origin of the confined carriers is supposed to be different, i.e. oxygen vacancies in GAO/STO and a polar discontinuity in LAO/STO. The most significant differences concern the mobility of the charge carriers, which is significantly higher in GAO/STO, and the thickness of the depth profile of the confined electrons which is smaller. The stronger carrier confinement in the GAO/STO heterostructure is also impressive, since unlike LAO/STO, it is not subjected to any post annealing procedure. Its higher electron mobility can be understood in terms of reduced disorder and strain effects at the GAO/STO interface, while the stronger carrier confinement may be the result of a stronger confining potential.

\begin{acknowledgments}
The work at the University of Fribourg has been supported by the Schweizerische Nationalfonds (SNF) through the grants No. 200020-153660. The work at Muni was supported by the project CEITEC(CZ.1.05/1.1.00/02.0068). Funding from the Danish Agency for Science, Technology and Innovation  (4070-00047B) is also acknowledged.
\end{acknowledgments}

\break

\begin{figure*}
\vspace*{0cm}
\hspace*{0cm}
\includegraphics[width=\textwidth]{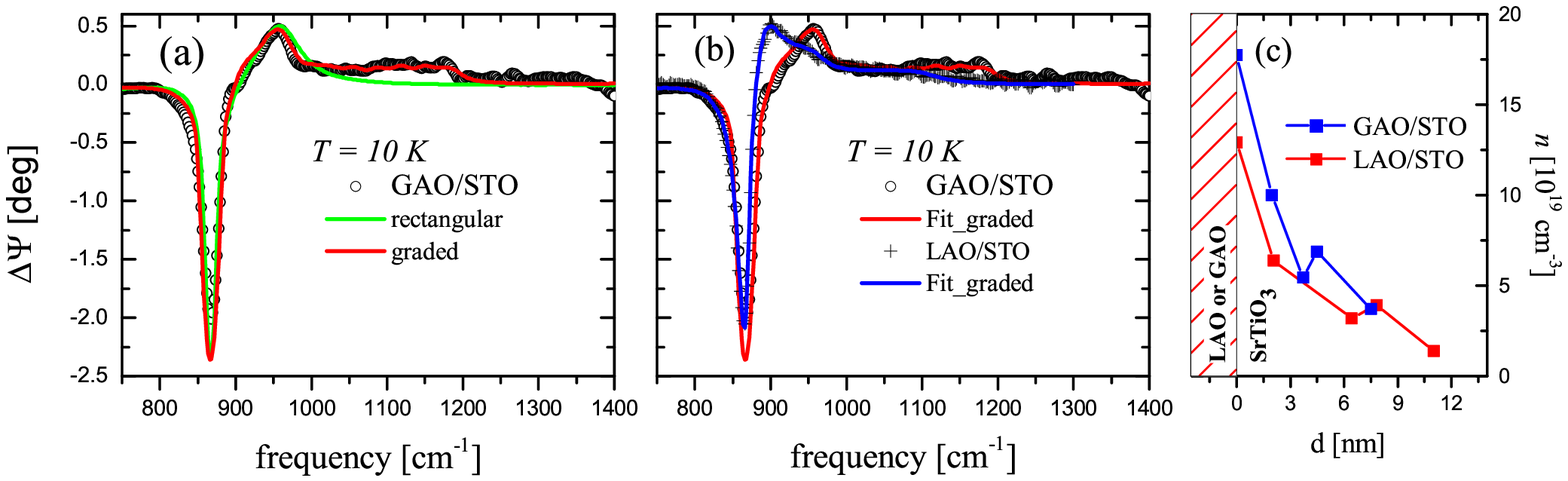}
\vspace{0cm}

\caption{\label{fig1SchemeSetup}
Difference spectrum of the ellipsometric angle, $\Delta\Psi=\Psi\mathrm{(GAO/STO)-\Psi(STO)}$, showing the Berreman mode above the highest LO phonon mode of STO. The experimental data (open symbols) are compared with the fits with an isotropic model using a rectangular carrier concentration profile (green line) and a graded profile (red line), respectively. (b) Comparison with the experimental data on LAO/STO (symbols) and the best fit with a graded profile (blue solid line) as reproduced from Ref.~\cite{Dubroka2010}. (c) Comparison of the depth profile of the carrier concentration, \textit{n}, as obtained from the fits with the graded potential.}
\end{figure*}

\begin{table}
\caption{\label{tab:table1}Parameters obtained from the best fit with an isotropic Drude response and a graded potential of the confined charge carriers at 10K. The parameters for LAO/STO have been reproduced from Ref.~\cite{Dubroka2010}.}

\begin{ruledtabular}
\begin{tabular}{c|rrc}
 &\textit{$N_{s}$} (cm$^{-2}$) &\textit{$\mu$} (cm$^{2}$/Vs) & \textit{d} (nm)\\
\hline
LAO/STO & $5.4\times10^{13}$ & 34 & 11\\
GAO/STO & $6.2\times10^{13}$ & 74 & 7.5\\

\end{tabular}
\end{ruledtabular}
\end{table}

\cleardoublepage

\begin{figure*}
\vspace*{0cm}
\hspace*{0cm}
\includegraphics[width=\textwidth]{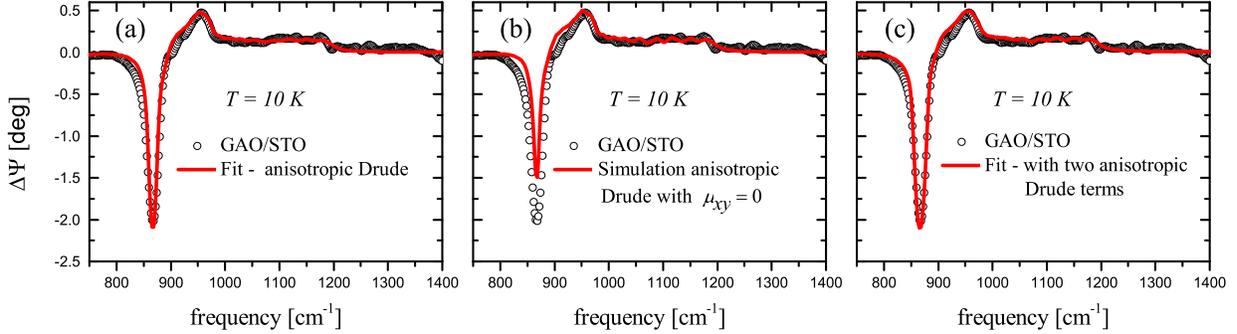}
\vspace{0cm}

\caption{\label{fig2STO_85nm}
Fitting of the difference spectrum $\Delta\Psi=\Psi\mathrm{(GAO/STO)-\Psi(STO)}$ with anisotropic Drude models. (a) Fit using a single anisotropic Drude term with $\mu_\textit{xy}$ as fit parameter and the remaining parameters fixed to the values obtained from the isotropic fit as shown in Figure 1 and Table 1. (b) Simulation of only the contribution of the out-of-plane component, i.e. for $\mu_\textit{xy}=0$. (c) Fit using two unidirectional Drude-terms, one for the out-of-plane reaspone with $\mu_\textit{z}=74$ and $\mu_\textit{xy}=0$ and the other for the in-plane response with $\mu_\textit{z}=0$ and $\mu_\textit{xy}>0$. 
}
\end{figure*}

\begin{table}
\caption{\label{tab:table1}Parameters obtained from the best fit using an anisotropic Drude response of the confined charge carriers along the out-of-plane (\textit{z}) and the in-plane (\textit{xy}) directions. Details are described in the text. }
\begin{ruledtabular}
\begin{tabular}{c|c|c|c}
 &\textit{$N_{s}$} (cm$^{-2}$) for \textit{z} direction&\textit{$\mu$} (cm$^{2}$/Vs) for \textit{xy} direction & \textit{$\mu_{xy}$} (cm$^{2}$/Vs)\\
\hline
Single component model & $6.2\times10^{13}$ & - & 5\\
Two component model & $6.2\times10^{13}$ & $4.6\times10^{13}$ & $>30$\\

\end{tabular}
\end{ruledtabular}
\end{table}


\begin{thebibliography}{99} \markboth{Bibliography}{Bibliography}

\bibitem{Mannhart2008}
J. Mannhart \etal, MRS Bulletin {\bf 33}, 1027-1034 (2008).

\bibitem{Cen2009}
C. Cen \etal, Science {\bf 323}, 1026-1030 (2009).

\bibitem{Caviglia2008}
A. D. Caviglia \etal, Nature {\bf 456}, 624-627 (2008).

\bibitem{Willmott2007}
P. R. Willmott \etal, Phys. Rev. Lett. {\bf 99}, 155502	(2007).

\bibitem{Kalabukhov2007}
A. Kalabukhov \etal, Phys. Rev. B {\bf 75}, 121404 (2007).

\bibitem{Herranz2012}
G. Herranz \etal, Scientific  Reports {\bf 2}, 758 (2012).

\bibitem{Chen2011}
Y. Z. Chen \etal, Nano Lett.  {\bf 11}, 3774-3778   (2011).

\bibitem{Lee2012}
S. W. Lee \etal, Nano Lett.  {\bf 12}, 4775-4783  (2012).

\bibitem{Chen2013}
Y. Z. Chen \etal, Nature Comm.  {\bf 4}, 1371  (2013).

\bibitem{Kormondy2015}
K. J. Kormondy \etal, J. Appl. Phys.  {\bf 117}, 095303 (2015).

\bibitem{Liu 2013}
Z. Q. Liu \etal, Phys. Rev. X  {\bf 3}, 021010  (2013).

\bibitem{Schutz2015}
P. Sch\"utz \etal, Phys. Rev. B  {\bf 91}, 165118 (2015).

\bibitem{Dubroka2010}
A. Dubroka \etal, Phys. Rev. Lett. {\bf 104}, 156807 (2010).

\bibitem{Bernhard2004}
C. Bernhard \etal, Thin Solid Films {\bf 455}, 143-149 (2004). 

\bibitem{Kozuka2007}
Y. Kozuka \etal, Phys. Rev. B {\bf 76}, 085129 (2007).

\bibitem{Berreman1963}
D. W. Berreman \etal, Phys. Rev. {\bf 130}, 2193-2198 (1963).

\bibitem{Humlíček1996}
J. Humli\v{c}ek \etal, Appl. Phys. Lett. {\bf 69}, 2581-2583 (1996).

\bibitem{Rossle2013}
M. R\"ossle \etal, Phys. Rev. B {\bf 88}, 104110 (2013).

\bibitem{Skoromets2014}
S. Y. Park \etal, Phys. Rev. B {\bf 87}, 205145 (2013).

\bibitem{Sing2009}
M. Sing \etal, Phys. Rev. Lett. {\bf 102}, 176805 (2009).

\bibitem{Basletic2008}
M. Basletic \etal, Nature Mat. {\bf 7}, 621-625 (2008).

\bibitem{van Mechelen2008}
J. L. M. van Mechelen \etal, Phys. Rev. Lett. {\bf 100}, 226403 (2008).



\end{thebibliography}
\end{document}